\documentclass{webofc}
\usepackage[varg]{txfonts}   % Web of Conferences font
\usepackage{hyperref}
\usepackage{url}
\usepackage{amsmath,bm}
\hypersetup{colorlinks=true,citecolor=blue,urlcolor=blue,linkcolor=blue}

\newcommand{\be}{\begin{equation}}
\newcommand{\ee}{\end{equation}}
\newcommand{\ba}{\begin{eqnarray}}
\newcommand{\ea}{\end{eqnarray}}
\newcommand{\bd}{\begin{displaymath}}
\newcommand{\ed}{\end{displaymath}}

\def\thalf{{\textstyle{\frac{1}{2}}}}

\begin{document}
\title{Relativistic hydrodynamics with spinodal decomposition}

\author{\firstname{Joseph} \lastname{Kapusta}\inst{1} \and
        \firstname{Mayank} \lastname{Singh}\inst{2}\fnsep\thanks{\email{mayank.singh@vanderbilt.edu}} \and
        \firstname{Thomas} \lastname{Welle}\inst{3}
        % etc.
}

\institute{School of Physics and Astronomy, University of Minnesota, Minneapolis, MN 55455, USA
\and
           Department of Physics and Astronomy, Vanderbilt University, Nashville, TN 37240, USA
\and
           Applied Research Associated, 8537 Six Forks Road, Raleigh, NC 27615, USA
          }

\abstract{We introduce the equations of relativistic hydrodynamics that incorporate phase separation via spinodal decomposition. These equations consider surface effects between the two phases and are applicable for simulating intermediate-energy heavy-ion collisions and binary neutron star mergers, where a first-order phase transition is expected. We solve these equations in the context of Bjorken flow, which offers the relevant geometric framework for ion collisions.
}
\maketitle
\section{Introduction}
Relativistic hydrodynamics serves as an effective theory for describing the dynamics of strongly interacting nuclear matter in ion collider experiments and astrophysical phenomena. The nuclear physics input in these hydrodynamic models comes from the equation of state of quantum chromodynamics (QCD). The QCD equation of state at zero baryon chemical potential is calculated from first principles using lattice gauge theory \cite{Aoki:2006we, Ding:2015ona}. The zero baryon chemical potential QCD equation of state is applicable for the highest energy nuclear collisions.

Recently, the Relativistic Heavy-Ion Collider (RHIC) completed the Beam Energy Scan (BES), which studied ion collisions at lower energies to investigate the finite baryon density region of the QCD phase diagram. Upcoming experiments at the Facility for Antiproton and Ion Research (FAIR) will further this research by probing the higher baryon chemical potential areas of the QCD phase diagram. Additionally, even greater chemical potentials are explored within the interiors of neutron stars and during binary neutron star mergers.

The QCD phase diagram is believed to exhibit a first-order phase transition at large baryon chemical potential, with this transition curve expected to terminate at a critical point \cite{Fukushima:2010bq}. This curve may be accessible through intermediate energy ion collision experiments or binary neutron star mergers. Therefore, understanding the dynamical signatures of this transition in observables is vital. Phase transition in nuclear collisions has been previously studied \cite{Randrup:2009gp, Randrup:2010ax}.This paper presents a covariant hydrodynamic framework that includes spinodal decomposition to study dynamical phase separation. \cite{Kapusta:2024nii}. We further solve the equation in Bjorken geometry to explore its implications for ion collisions.

\section{Spinodal decomposition in relativistic hydrodynamics}

At constant temperature $T$ and baryon chemical potential $n({\bf x}, t)$, we can consider the Helmholtz free energy $F$ 

\begin{equation}
F\{n({\bf x},t)\} = \int d^3x \left[ \thalf K (\bm{\nabla} n)^2 + f(T,n) \right],
\end{equation}
where $f(T,n)$ is the bulk free energy. The $K$ is the coefficient of the surface energy, and the first term accounts for energy due to the phase boundary.

In a non-relativistic setting, this leads to the stress-energy tensor \cite{CahnHilliard1, CahnHilliard2}
\begin{equation}
T_{ij} = \tilde{P} \delta_{ij} + K (\partial_i n) (\partial_j n),
\end{equation}
where $\tilde{P}$ is the local thermodynamic pressure. The relativistic generalization of this can be written as

\begin{equation}
T^{\mu\nu} = \tilde{P} (u^{\mu} u^{\nu} - g^{\mu\nu}) + \tilde{\epsilon} u^{\mu} u^{\nu} + K (D^{\mu} n) (D^{\nu} n).
\end{equation}
We are using a mostly negative metric $g^{\mu\nu} = {\rm diag.} (+---)$. The fluid four-velocity is given by $u^\mu$ with the normalization $u^\mu u_\mu = 1$. Local thermodynamic pressure $\tilde{P}$ and energy density $\tilde{\epsilon}$ have contributions from the surface terms.
\be
\tilde{P} = P + KnD^2n + \frac{1}{2}K(D^\mu n)(D_\mu n),
\ee
\be
\tilde{\epsilon} = \epsilon - \frac{1}{2}K(D^\mu n)(D_\mu n),
\ee
where $D^\mu = \partial^\mu - u^\mu u^\alpha\partial_\alpha$ is the gradient orthogonal to the fluid velocity.

The baryon current $J^\mu$ is given as
\begin{equation}
J^{\mu} = n u^{\mu} + \sigma_B T D^{\mu} \left( \frac{\tilde{\mu}}{T} \right),
\end{equation}
where $\sigma_B$ is the baryon conductivity. Baryon chemical potential $\tilde{\mu}$ also has contributions from surface terms
\begin{equation}
    \tilde{\mu} = \mu + KD^2n
\end{equation}

\begin{figure}
    \centering
    \includegraphics[width=0.5\linewidth]{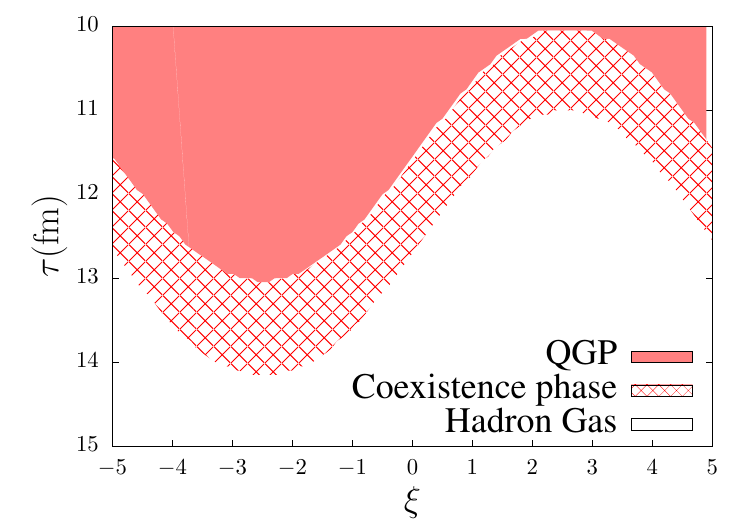}
    \caption{System phase at different spacetime points.}
    \label{fig:fig1}
\end{figure}

\section{Bjorken flow}

\begin{figure}
    \centering
    \includegraphics[width=0.5\linewidth]{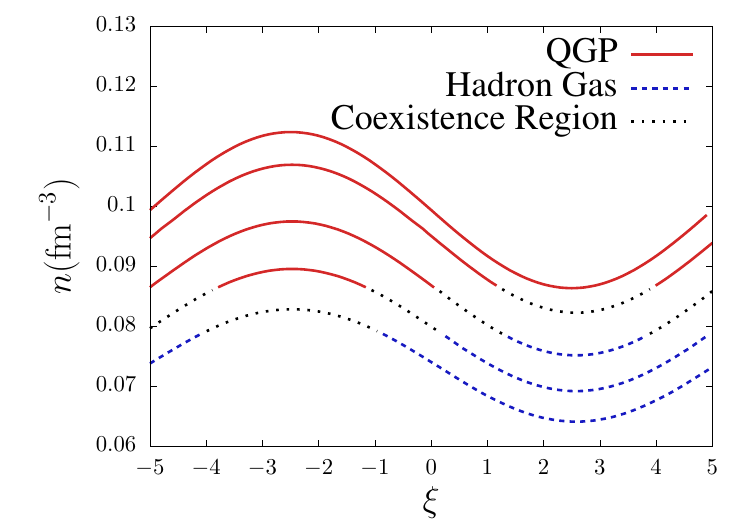}
    \caption{Baryon density evolution at different times. From top to bottom, curves correspond to $\tau =10$ fm, 10.5 fm, 11.5 fm, 12.5 fm and 13.5 fm.}
    \label{fig:fig2}
\end{figure}

The rapid longitudinal expansion in heavy-ion collisions can be approximated by Bjorken flow \cite{Cooper:1974qi, Chiu:1975hw, Bjorken:1982qr}. Using the longitudinal proper time $\tau = \sqrt{t^2 - z^2}$ and the spacetime rapidity $\xi = \tanh^{-1}(z/t)$, the energy-momentum and baryon current conservation equations can be reduced to the following system of equations:
\be\label{eq:solve1}
\frac{\partial \epsilon(n,T)}{\partial \tau} + \frac{w(n,T)}{\tau} + \frac{K}{\tau^2} \frac{\partial n}{\partial \xi}  \frac{\partial^2 n}{\partial \tau \partial \xi} 
- \frac{K}{\tau^3} n \frac{\partial^2 n}{\partial \xi^2} = 0,
\ee
\be\label{eq:solve2}
\frac{\partial}{\partial \tau} (\tau n) - \frac{\sigma_B T}{\tau} \frac{\partial^2}{\partial \xi^2} \left( \frac{\tilde{\mu}}{T} \right)
- \frac{1}{\tau} \frac{\partial}{\partial \xi} (\sigma_B T) \frac{\partial}{\partial \xi} \left( \frac{\tilde{\mu}}{T} \right) = 0,
\ee
\be\label{eq:solve3}
\tilde{\mu} = \mu(n,T) - \frac{K}{\tau^2} \frac{\partial^2 n}{\partial \xi^2}.
\ee

Additionally, an equation of state is required to close the system. We use the background equation of state constructed by matching the equations of state from perturbative QCD to those from the hadron resonance gas, utilizing a switching function for the transition \cite{Albright:2014gva}. The free parameters in the equation of state are constrained by matching to lattice QCD results at zero baryon chemical potential. A critical point with a first order phase transition curve was embedded on the background equation of state \cite{Kapusta:2021oco}. The equation of state further needs to be interpolated to the metastable and unstable regions. We do that following the prescription shown in \cite{Kapusta:2024nii}.

The equations are evolved with the simplification that the energy density is assumed to be boost invariant. Effectively, we neglect the energy density gradients. The violation in energy density boost invariance was found to be less than 3\%. We start with a constant energy density and a sinusoidal baryon density distribution in rapidity. The baryon density is chosen such that the lowest baryon density point just touches the phase transition curve on the phase diagram. The start time is $\tau = 10$ fm. We choose $K = 0$ and $K = 5\times10^{-5}$ MeV$^{-4}$ to highlight the effect of surface energy terms.

\begin{figure}
    \centering
    \includegraphics[width=0.45\linewidth]{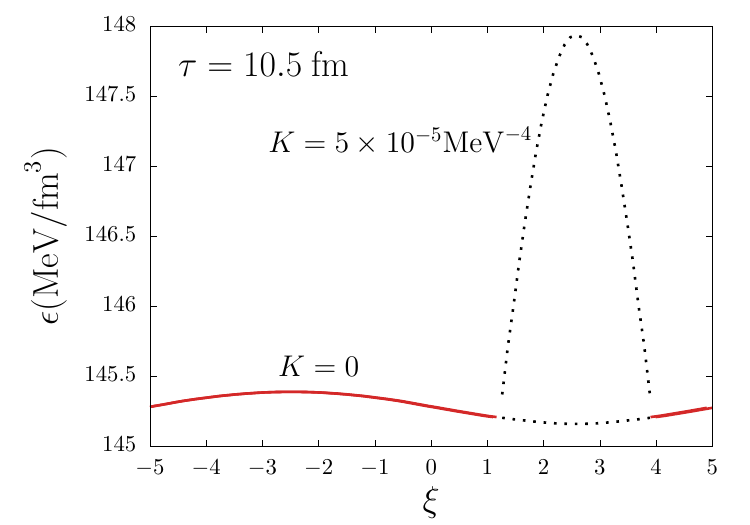}
    \includegraphics[width=0.45\linewidth]{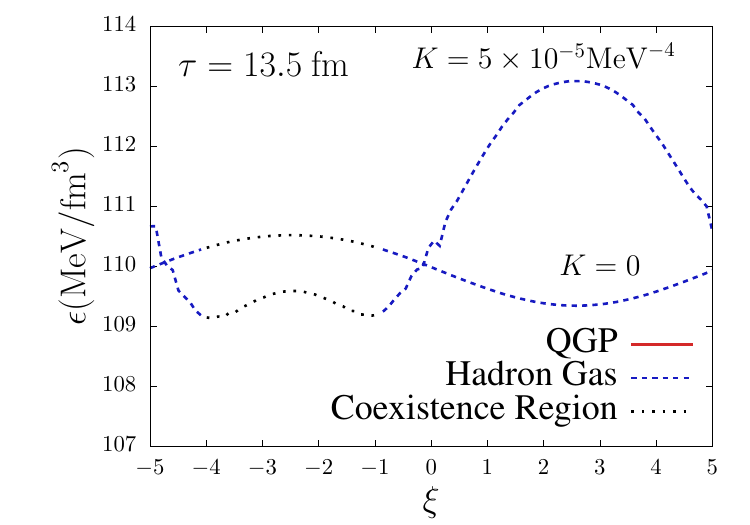}
    \caption{Energy density evolution at two different values of $\tau$.}
    \label{fig:fig3}
\end{figure}

The QCD phases at different spacetime points are shown in Fig. \ref{fig:fig1} while Fig. \ref{fig:fig2} depicts the evolution of baryon density. The system transitions from an entirely quark-gluon plasma (QGP) phase to a complete hadron gas phase as a result of rapid expansion. During this process, it traverses the coexistence region, which leaves a lasting imprint. Figure \ref{fig:fig3} shows energy density at $\tau = 10.5$ fm and $\tau = 13.5$ fm.The energy density varies as the system crosses the crossover region, influenced by the baryon density gradient; it may be either higher or lower. This variation continues even after the system fully transitions to the hadron gas phase.

\section{Summary}

We present covariant equations of relativistic hydrodynamics including spinodal decomposition. By addressing surface effects between distinct phases, these equations are particularly relevant for simulating intermediate-energy heavy-ion collisions and binary neutron star mergers, where first-order phase transitions are expected. We solved the resulting equations within Bjorken flow, which captures the relevant geometry for ion collisions. 

\section*{Acknowledgements}

This work was supported by the U.S. Department of Energy Grant Nos. DE-FG02-87ER40328 (JK, MS and TW) and DE-SC-0024347 (MS).


\begin{thebibliography}{}
%\cite{Aoki:2006we}
\bibitem{Aoki:2006we}
Y.~Aoki, G.~Endrodi, Z.~Fodor, S.~D.~Katz and K.~K.~Szabo,
%``The Order of the quantum chromodynamics transition predicted by the standard model of particle physics,''
Nature \textbf{443}, 675-678 (2006)
%doi:10.1038/nature05120
%[arXiv:hep-lat/0611014 [hep-lat]].
%1968 citations counted in INSPIRE as of 29 May 2024

%\cite{Ding:2015ona}
\bibitem{Ding:2015ona}
H.~T.~Ding, F.~Karsch and S.~Mukherjee,
%``Thermodynamics of strong-interaction matter from Lattice QCD,''
Int. J. Mod. Phys. E \textbf{24}, no.10, 1530007 (2015)
%doi:10.1142/S0218301315300076
%[arXiv:1504.05274 [hep-lat]].
%426 citations counted in INSPIRE as of 30 Sep 2025

%\cite{Fukushima:2010bq}
\bibitem{Fukushima:2010bq}
K.~Fukushima and T.~Hatsuda,
%``The phase diagram of dense QCD,''
Rept. Prog. Phys. \textbf{74}, 014001 (2011)
%doi:10.1088/0034-4885/74/1/014001
%[arXiv:1005.4814 [hep-ph]].
%934 citations counted in INSPIRE as of 29 May 2024

%\cite{Randrup:2009gp}
\bibitem{Randrup:2009gp}
J.~Randrup,
%``Phase transition dynamics for baryon-dense matter,''
Phys. Rev. C \textbf{79}, 054911 (2009)
%doi:10.1103/PhysRevC.79.054911
%[arXiv:0903.4736 [nucl-th]].
%108 citations counted in INSPIRE as of 30 Sep 2025

%\cite{Randrup:2010ax}
\bibitem{Randrup:2010ax}
J.~Randrup,
%``Spinodal phase separation in relativistic nuclear collisions,''
Phys. Rev. C \textbf{82}, 034902 (2010)
%doi:10.1103/PhysRevC.82.034902
%[arXiv:1007.1448 [nucl-th]].
%70 citations counted in INSPIRE as of 30 Sep 2025
%\cite{Kapusta:2024nii}
\bibitem{Kapusta:2024nii}
J.~I.~Kapusta, M.~Singh and T.~Welle,
%``Covariant formulation of spinodal decomposition in rapidly expanding quark gluon plasma,''
Phys. Rev. C \textbf{110}, no.5, 054902 (2024)
%doi:10.1103/PhysRevC.110.054902
%[arXiv:2407.16963 [hep-ph]].
%2 citations counted in INSPIRE as of 03 Dec 2024

\bibitem{CahnHilliard1}
J.~W.~Cahn and J.~E.~Hilliard,
%``Free Energy of a Nonuniform System. I. Interfacial Free Energy,''
J. Chem. Phys. \textbf{28}, 258 (1958)
%doi:10.1063/1.1744102

\bibitem{CahnHilliard2}
J.~W.~Cahn and J.~E.~Hilliard,
%``Free Energy of a Nonuniform System. III. Nucleation in a Two‐Component Incompressible Fluid,''
J. Chem. Phys. \textbf{31}, 688 (1959)
%doi:10.1063/1.1730447
% etc

%\cite{Cooper:1974qi}
\bibitem{Cooper:1974qi}
F.~Cooper, G.~Frye and E.~Schonberg,
%``Landau's Hydrodynamic Model of Particle Production and electron Positron Annihilation Into Hadrons,''
Phys. Rev. D \textbf{11}, 192 (1975)
%doi:10.1103/PhysRevD.11.192
%204 citations counted in INSPIRE as of 30 Sep 2025

%\cite{Chiu:1975hw}
\bibitem{Chiu:1975hw}
C.~B.~Chiu, E.~C.~G.~Sudarshan and K.~H.~Wang,
%``Hydrodynamical Expansion with Frame Independence Symmetry in High-Energy Multiparticle Production,''
Phys. Rev. D \textbf{12}, 902 (1975)
%doi:10.1103/PhysRevD.12.902
%43 citations counted in INSPIRE as of 30 Sep 2025

%\cite{Bjorken:1982qr}
\bibitem{Bjorken:1982qr}
J.~D.~Bjorken,
%``Highly Relativistic Nucleus-Nucleus Collisions: The Central Rapidity Region,''
Phys. Rev. D \textbf{27}, 140-151 (1983)
%doi:10.1103/PhysRevD.27.140
%3798 citations counted in INSPIRE as of 30 Sep 2025

%\cite{Albright:2014gva}
\bibitem{Albright:2014gva}
M.~Albright, J.~Kapusta and C.~Young,
%``Matching Excluded Volume Hadron Resonance Gas Models and Perturbative QCD to Lattice Calculations,''
Phys. Rev. C \textbf{90}, no.2, 024915 (2014)
%doi:10.1103/PhysRevC.90.024915
%[arXiv:1404.7540 [nucl-th]].
%119 citations counted in INSPIRE as of 29 Sep 2025

%\cite{Kapusta:2021oco}
\bibitem{Kapusta:2021oco}
J.~Kapusta, T.~Welle and C.~Plumberg,
%``Embedding a Critical Point in a Hadron to Quark\textendash{}Gluon Crossover Equation of State,''
Phys. Rev. C \textbf{106}, no.1, 014909 (2022)
%doi:10.5506/APhysPolBSupp.16.1-A46
%[arXiv:2112.07563 [nucl-th]].
%6 citations counted in INSPIRE as of 29 May 2024


\end{thebibliography}
\end{document}